\documentclass[a4paper]{article}

\usepackage{INTERSPEECH2022}
\usepackage{multirow}
\usepackage{hyperref}

\author{Muyang Du\inst{1} \and
Chuan Liu\inst{1} \and
Jiaxing Qi\inst{2} \and
Junjie Lai\inst{2}}

\title{Efficient Incremental Text-to-Speech on GPUs}
\name{Muyang Du$^1$, Chuan Liu$^1$, Jiaxing Qi$^1$, Junjie Lai$^1$}
\address{
  $^1$NVIDIA Corporation}
\email{\{myrond,riverl,jqi,julienl\}@nvidia.com}

\begin{document}

\maketitle
\begin{abstract}

Incremental text-to-speech, also known as streaming TTS, has been increasingly applied to online speech applications that require ultra-low response latency to provide an optimal user experience. However, most of the existing speech synthesis pipelines deployed on GPU are still non-incremental, which uncovers limitations in high-concurrency scenarios, especially when the pipeline is built with end-to-end neural network models. To address this issue, we present a highly efficient approach to perform real-time incremental TTS on GPUs with Instant Request Pooling and Module-wise Dynamic Batching. Experimental results demonstrate that the proposed method is capable of producing high-quality speech with a first-chunk latency lower than 80ms under 100 QPS on a single NVIDIA A10 GPU and significantly outperforms the non-incremental twin in both concurrency and latency. Our work reveals the effectiveness of high-performance incremental TTS on GPUs.

  
\end{abstract}
\noindent\textbf{Index Terms}: incremental speech synthesis, streaming tts, neural networks, deep learning.

\section{Introduction}

\begin{figure*}[t]
  \centering
  \includegraphics[width=14.5cm]{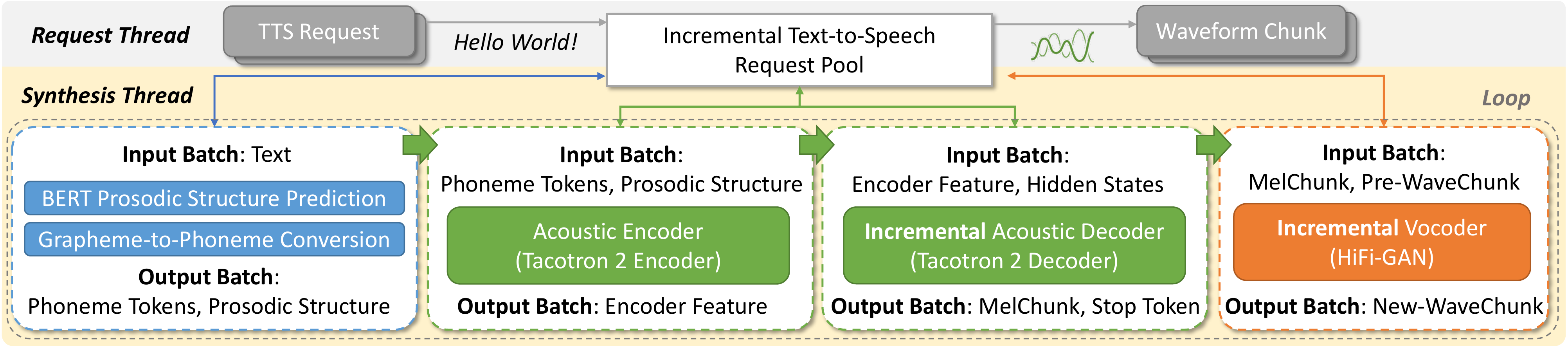}
  \caption{Architecture of the Proposed Efficient Incremental TTS Pipeline on GPU.}
  \label{fig:proposed_pipeline}
\vspace{-4mm}
\end{figure*}

With the recent blossoming in deep learning, speech synthesis methods have switched from traditional concatenation-based\cite{lee1993improved} and HMM-based\cite{yamagishi2009robust,pouget2015hmm} statistical parametric methods to neural network based methods and have been widely used in various application scenarios. Compared with traditional methods, neural networks can produce more natural and high-fidelity speech at the cost of more computing power and larger latency. Therefore, reducing the latency of the speech synthesis is vital to improve the user experience of applications that require instant response, such as the virtual agents of call centers, especially in the case of highly concurrent requests during peak periods. With such needs, incremental synthesis shows its benefits. Instead of synthesizing the entire speech audio before responding, incremental synthesis can produce speech chunk-by-chunk to provide a lower response latency. Once the first audio chunk is generated, the synthesis time of the subsequent audio chunks is hidden within the playback time of the preceding chunks.

Speech synthesis pipelines deployed in production environments typically comprises three major components: a frontend for extracting linguistic features from the text, an acoustic model for synthesizing acoustic features such as the Mel spectrogram, and a vocoder for converting acoustic features into waveform samples. The frontend typically performs text normalization, grapheme-to-phoneme\cite{hunnicutt1980grapheme} conversion, prosodic structure prediction, and named entity recognition. BERT\cite{devlin2018bert}, a transformer-based pre-training language model for natural language processing, has been applied in several unified TTS frontend research efforts\cite{zhang2020unified,bai2021universal} and has been proven capable of effectively extracting a wide range of linguistic features for realistic speech synthesis. Acoustic models can be grouped into two major types: autoregressive and parallel. Among them, the autoregressive model Tacotron2\cite{wang2017tacotron,shen2018natural} and its variants have dominated the acoustic models used in industry for many years due to their outstanding and stable synthesis quality. In recent years, parallel models, such as FastSpeech\cite{ren2019fastspeech,ren2020fastspeech} and FastPitch\cite{lancucki2021fastpitch}, have dramatically improved the controllability of speech, and are widely in personalized synthesis and singing synthesis. Parallel speech synthesis models often have architecture based on Transformer\cite{vaswani2017attention} and Convolution. They have higher throughput compared with autoregressive models for non-incremental synthesis. However, the synthetic speech quality of parallel models is highly dependent on the visibility of the entire feature sequence. In the case of incremental synthesis, the perceptual field is usually limited to the size of the chunk resulting in quality degradation of parallel models. In comparison, the hidden state of autoregressive models passed over time carrying contextual information ensures speech quality unharmed. Meanwhile, Samsung's study\cite{ellinas2021high} also pointed out that incremental synthesis using an autoregressive model has a more stable latency when the text length increases, which is evident as the latency is positively correlated with the computational overhead. 

Similar to acoustic models, there are autoregressive and non-autoregressive vocoders. Autoregressive vocoder models such as WaveNet\cite{oord2016wavenet}, WaveRNN\cite{kalchbrenner2018efficient}, and LPCNet\cite{valin2019lpcnet} synthesize audio sample-by-sample based on the acoustic features and previously generated samples. Parallel vocoders such as Flow-based WaveGlow\cite{prenger2019waveglow}, GAN-based MelGAN\cite{kumar2019melgan}, and HiFi-GAN\cite{kong2020hifi} directly map the entire acoustic feature segment to a waveform segment. Unlike acoustic models, autoregressive vocoders needs to perform an enormous number of regressions to generate a chunk of audio. Although the number of regressions can be reduced using the multi-band mechanism\cite{nguyen1994near}, further model compression may still be required to obtain real-time. In contrast, the parallel vocoder can perform chunk-by-chunk generation, which is friendly to GPUs and greatly accelerates both incremental and non-incremental synthesis.

With more applications relying on low latency speech synthesis in recent years, several studies\cite{stephenson2021alternate,mohan2020incremental,ma2019incremental, stephenson2020future} have been proposed to optimize the quality and naturalness of incremental TTS. However, previous studies have disregarded the performance of incremental TTS on GPUs, which should be worthy of attention, especially under high concurrency scenarios. To the best of our knowledge, many incremental TTS pipelines deployed in production are CPU-based, which typically requires reducing the computational cost by compressing\cite{cheng2017survey}, sparsifying\cite{hoefler2021sparsity}, and distilling\cite{hinton2015distilling} the model to obtain real-time synthesis. During inference, each TTS request is processed by several CPU cores, and no batching is involved. The additional optimization efforts for real-time on the CPU are usually time-consuming and may negatively impact the synthetic speech quality. Furthermore, the concurrency that each CPU server can handle is strictly limited by the number of CPU cores. In comparison, GPUs are better suited to processing computationally-intensive tasks with high-concurrency. Despite the fact that GPUs have been widely used in training and non-incremental synthesis as they are naturally friendly to batching, more needs to be investigated about efficient incremental TTS on GPUs.

To solve the above problems, we propose an efficient approach for incremental text-to-speech synthesis on GPUs. The proposal contains a BERT-based frontend, a Tacotron 2-based acoustic model, and a HiFi-GAN-based vocoder. Furthermore, we propose the use of Instant Request Pooling and Module-wise Dynamic Batching strategies, which are vital for highly efficient incremental synthesis on GPUs. In the experiments, we make pressure and listening tests on the pipeline running on NVIDIA A10 GPU. Experiments prove that our approach can produce high-fidelity speech with ultra-low latency under high QPS.

\section{Methods}

\subsection{Modules}

\subsubsection{Frontend}

The frontend comprises a grapheme-to-phoneme (G2P) conversion unit and a BERT-based prosodic structure prediction model. G2P first converts the text into a grapheme sequence using the forward maximum matching algorithm based on a pronunciation dict of Chinese phrases, then converts the pinyin of each Chinese character into its phoneme tokens based on a mapping table. During this process, the number of phoneme tokens for each Chinese character $char_{i}$ is also recorded as $count_{i}$. In order to predict the prosodic structure, the text is first passed through a shared BERT backbone to extract hidden prosodic features $h_{i}$ with multi-head attention blocks. Then three separate linear layers are used to predict the three-level prosodic structure tokens $pw$ (prosodic word), $pph$ (phonological phrase), and $iph$ (intonational phrase). Finally, the prosodic structure token sequence are regulated using $count_{i}$ to have the same length as the phoneme token sequence.
\subsubsection{Acoustic Encoder}

The acoustic encoder is based on the Tacotron 2 encoder. It contains a token embedding layer, three stacked convolutional layers with batch regularization and ReLU activation, and a bi-directional LSTM layer. For simplicity, we refer to it as the CBRL network. In addition, three embedding layers are introduced to encode the three-level prosodic structure information. The prosodic structure embeddings $E_{pw}$, $E_{pph}$, $E_{iph}$ are summed with the token embeddings $E_{token}$ and then passed to the CBRL network to generate the encoded feature $F_{enc}$.
\begin{equation}
\begin{aligned}
    E_{all} &= E_{token} + E_{pw} + E_{pph} + E_{iph}\\
    F_{enc} &= CBRL(E_{all}) 
\end{aligned}
\end{equation}
\subsubsection{Acoustic Decoder} 

The acoustic decoder is based on the Tacotron 2 decoder. It contains an information bottleneck PreNet, a location-sensitive attention module $LSA$, two stacked unidirectional LSTM layers, two linear layers for predicting Mel spectrogram and stop token, and a PostNet for improving the Mel reconstruction. Unlike the conventional Tacotron 2 decoder that generates the entire Mel spectrogram at once. Incremental synthesis generates only a chunk of Mel spectrogram frames at a time, meaning the decoder states carrying contextual information across chunks need to be explicitly maintained during the decoding process. These states include the last Mel frame $M_{last}$, the attention context $A$, the attention weight of the last Mel frame $W_{last}$, the accumulated attention weights $W_{acc}$, the hidden states $H_{att}$, $H_{dec}$ and the cell states $C_{att}$, $C_{dec}$ of the two LSTM layers.
\begin{equation}
\begin{aligned}
    H_{att}, C_{att} &= LSTM_{att}([Pre(M_{last}), A], H_{att}, C_{att})\\
    A, W_{last} &= LSA(H_{att}, F_{enc}, W_{acc})\\
    H_{dec}, C_{dec} &= LSTM_{dec}([H_{att}, A], H_{dec}, C_{dec})\\
    W_{acc} &= W_{acc} + W_{last}\\
    M_{last} &= Linear(H_{dec}, A)
\end{aligned}
\end{equation}

\subsubsection{Vocoder Model} 

\begin{figure*}[t]
  \centering
  \includegraphics[width=14.5cm]{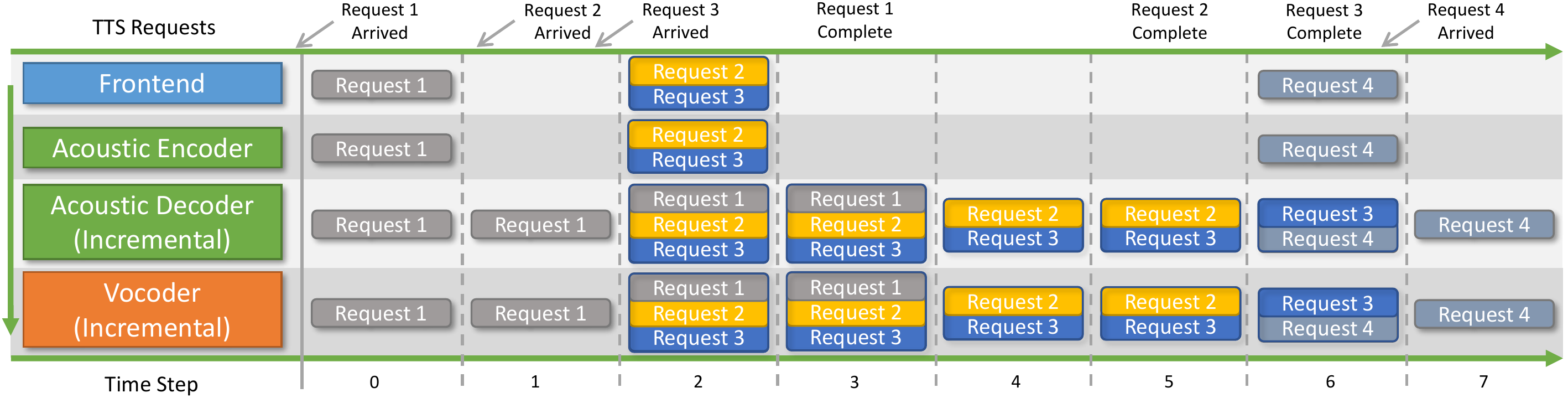}
  \caption{Incremental Synthesis Timeline with the Proposed Pooling and Batching Strategies.}
  \label{fig:inference_timeline}
\vspace{-4mm}
\end{figure*}

The vocoder is a HiFi-GAN-based generator $G$ that contains multiple transposed convolution and multi-receptive field fusion (MRF) blocks. In incremental synthesis, the vocoder synthesizes one waveform chunk at a time based on the Mel spectrogram chunk. In order to provide a smooth articulation between two adjacent waveform chunks, we splice multiple frames from the tail of the previous Mel spectrogram chunk $M_{pre}$ to the head of the current Mel spectrogram chunk $M_{cur}$ and then pass it through the vocoder. Finally, the overlap areas between the current audio chunk $S_{cur}$ and the previous audio chunk $S_{pre}$ are fused by applying fade-in and fade-out coefficients $\alpha$, $\beta$ from equal power cross-fade, as illustrated in Equation 3.
\begin{equation}
\begin{aligned}
    S_{cur} & = G([tail(M_{pre}), M_{cur}])\\
    head(S_{cur}) &= \alpha \odot head(S_{cur}) + 
\beta \odot tail(S_{pre})
\end{aligned}
\end{equation}

\subsection{Incremental Synthesis}


\subsubsection{Incremental Synthesis of a Single Text}

In order to achieve ultra-low latency, the proposed approach generates one short audio chunk at a time. In practice, each request contains a single text corresponding to multiple responses. Each response carries an audio chunk of several hundred milliseconds in duration, depending on the configurable size of the chunk. The synthesis of each request involves four modules, the frontend ($F$), the acoustic encoder ($E$), the acoustic decoder ($D$), and the vocoder ($V$). The frontend and acoustic encoder modules are non-incremental and run only once per text. The acoustic decoder and vocoder modules are incremental and run several times for each text, synthesizing one audio chunk at a time. The stop token output by the acoustic model decoder exceeding the specified threshold flags the end of synthesis.

\subsubsection{Incremental Synthesis under High Concurrency}

Incremental synthesis under high concurrency faces two challenges. Firstly, scheduling numerous requests while ensuring that each request is processed as soon as it reaches the server. Secondly, synthesizing multiple requests simultaneously while keeping low latency. As shown in Figure 1, we introduce two strategies, Instant Request Pooling and Module-wise Dynamic Batching, to guarantee new requests are synthesized instantly, together with all the other incomplete requests in the server.

\noindent$\blacksquare$ \textbf{Instant Request Pooling}. Conventional approach to handle the incoming TTS requests while the server is processing the ongoing requests is to keep the new requests waiting inside a queue to be processed when the ongoing requests are complete. However, this approach can hardly meet the requirement of low-latency incremental TTS because the waiting time can be long under high concurrency. Therefore, we introduce a request pool to hold all the ongoing and new requests. Once a new request arrives, it is immediately added to the pool as a pool item and available for synthesis. An infinite loop handles the incremental synthesis. New requests received during the current iteration only need to wait for the next iteration to be processed, regardless of the number of ongoing requests in the pool. The primary features saved in each pool item is described as follows:
\begin{itemize}
  \item [1.]
 Text and Model States, including frontend outputs phoneme and prosody tokens, encoder output $F_{enc}$, decoder states $M_{last}$, $W_{last}$, $A$, $W_{acc}$, $H_{att}$, $H_{dec}$, $C_{att}$, $C_{dec}$, and vocoder states $M_{pre}$, $M_{cur}$, $S_{pre}$, $S_{cur}$.
  \item [2.]
 Module Indicator, indicating which module should process this item in the current loop iteration. Initialized to 0 for $F$, $E$, then changed to 1 for $D$, $V$ when $E$ completes.
\end{itemize}
                                      
\noindent$\blacksquare$ \textbf{Module-wise Dynamic Batching}. Batching is an essential strategy to fully utilize GPU's parallel computing capability for both training and inference. For incremental TTS, we propose to use a dynamic batch size for each module. In each loop iteration, all the modules are executed sequentially, and the batch size of each module is determined by the number of items in the pool that each module should process. Specifically, each module first retrieves the pool items it should process based on the module indicator saved in each pool item. Then, it constructs an input batch for each model state from the retrieved items and performs a batched inference. Finally, it splits the output batch of each state and saves the updated state back to its corresponding pool item. Take the vocoder as an example, suppose $N$ pool items are retrieved for the vocoder in the current iteration. Their states $[M_{pre}^{1}, M_{pre}^{2}, ..., M_{pre}^{N}]$, $[M_{cur}^{1}, M_{cur}^{2}, ..., M_{cur}^{N}]$, and $[S_{pre}^{1}, S_{pre}^{2}, ..., S_{pre}^{N}]$ are constructed to input batch $M_{pre}^{B}$, $M_{cur}^{B}$, and $S_{pre}^{B}$. After inference, the output batch $S_{cur}^{B}$ is split into $[S_{cur}^{1}, S_{cur}^{2}, ..., S_{cur}^{N}]$ and we update each retrieved item $i$:
\begin{equation}
\begin{aligned}
    M_{pre}^{i} = M_{cur}^{i},\ S_{pre}^{i} = S_{cur}^{i}\\
\end{aligned}
\end{equation}
During synthesis, the decoder and vocoder are incremental and their states are constantly updated in each iteration. If a request completes, its item is removed from the pool immediately.

\subsubsection{Timeline and Latency Analysis}

Figure 2 elaborates on the timeline of incremental synthesis with the proposed pooling and batching strategies. Each time step corresponds to one incremental synthesis iteration. In short, a new request can be processed in step $t$ if it arrives before the start of the frontend module in step $t$. Request 1 arrives right before step 0 and is processed in step 0. Requests 2, 3 arrive during the execution of step 1 and are processed in step 2 in dynamic batches along with request 1. Request 1 is completed and removed immediately from the pool in step 3. Requests 2, 3 are processed together in dynamic batches in steps 4, 5. Request 2 is then completed and removed in step 5. Request 4 arrives at the end of step 5 and is processed in dynamic batches in step 6 with request 3, which is completed and removed in step 6.

Define the execution time of time step t as $T_t$. If a request arrives at step $t$, the maximum first audio chunk latency is $T_{t} + T_{t+1}$ (arrives right after the start of the frontend in step $t$), the minimum latency is $min(T_{t}, T_{t+1})$ (arrives right before the start of the frontend in step $t$, or right before the end of step $t$), and the expectation of the latency is $\mathbb{E} = T_{t}/2 + T_{t+1}$.

\begin{figure*}[t]
  \centering
  \includegraphics[width=15cm]{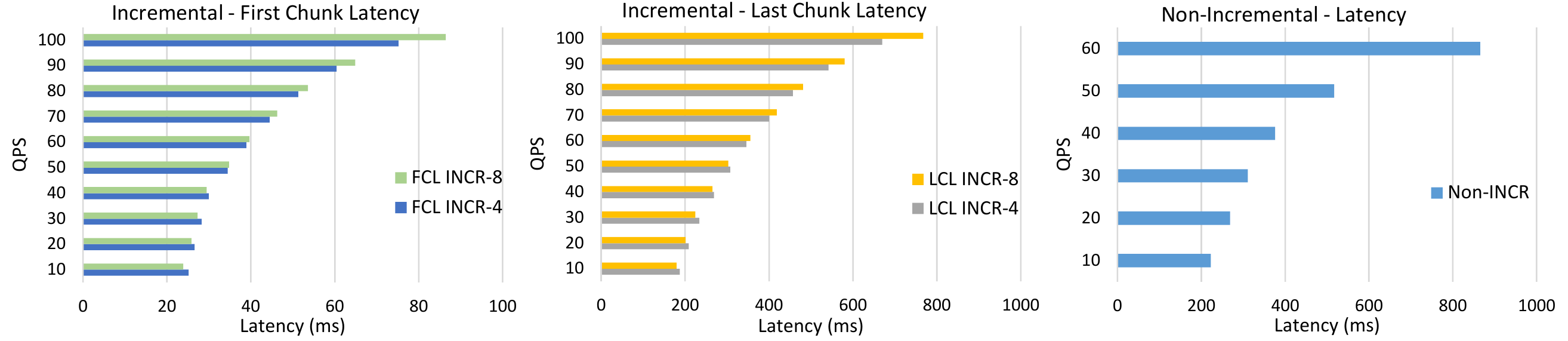}
  \caption{Response latency of the Mixed-Length Text under different QPS.}
  \label{fig:response_latency}
\vspace{-4mm}
\end{figure*}

\section{Experiments}

\begin{figure}[t]
  \centering
  \includegraphics[width=7cm]{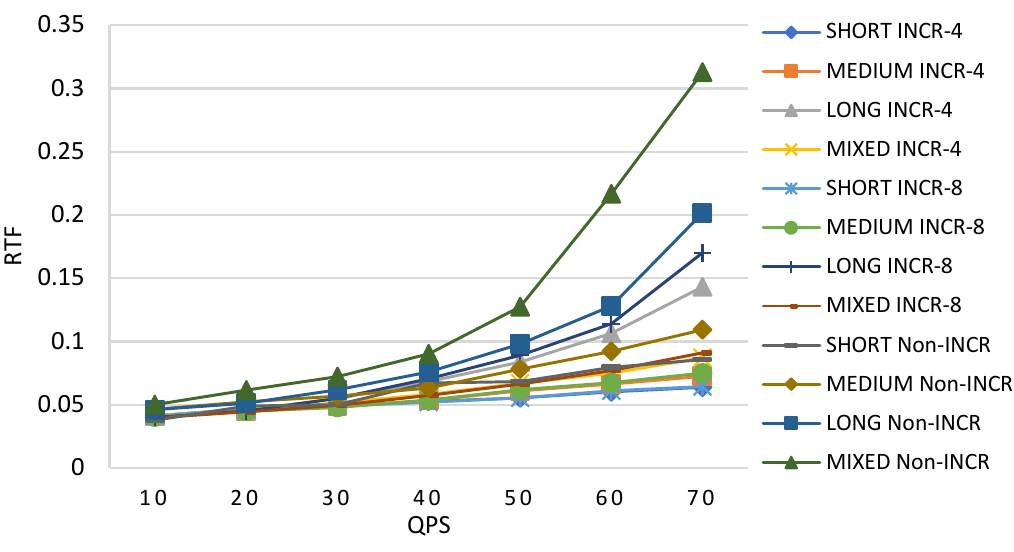}
  \caption{Real Time Factor under different QPS.}
  \label{fig:real_time_factor}
\vspace{-4mm}
\end{figure}

\subsection{Settings}

All the models in our experiments are trained with the Chinese Standard Mandarin Speech Corpus (CSMSC)\cite{databakercsmsc} open-sourced by Databaker. The Bert used in the frontend contains 12 Transformer layers with each layer 12 heads and 768 hidden units. The Tacotron 2 has the same model parameters as the original proposal with three additional 512-dim prosody embedding table. The HiFi-GAN vocoder with 128 upsample initial convolution channels and upsample factors of [8,8,2,2] is used to synthesize 22.05kHz 16bit PCM audio. All the models are trained on NVIDIA Tesla V100 GPU, and the synthesis pipelines are tested on NVIDIA Ampere A10 GPU. The proposed incremental pipeline is implemented in C++ as a NVIDIA Triton custom backend and the models are accelerated with TensorRT.

Since there is no publicly available performance benchmark for incremental TTS on GPU under high concurrency, we compare the proposed approach with a self-developed non-incremental pipeline using the exact same models. For the non-incremental twin, the TTS requests received over a short period are constructed into a batch for inference. The size of the batch keeps unchanged for all the models within each inference round. New requests received during the current round need to wait in the queue for the next round to be processed. Moreover, TensorRT and Triton are also applied for a fair comparison.

In the experiments, a fixed chunk size of 32 ($\sim$372ms) Mel frames is used with an overlap (OL) of 4 and 8. We test both the incremental and non-incremental pipelines with three lengths of text (TL) and their mixture. The average audio duration (DUR) for each text type are Short (2.64s), Medium (4.08s), Long (7.00s), and Mixed (4.54s). For incremental synthesis (INCR), we measure the first-chunk latency (FCL) (between the request being sent and the first chunk being received), the last-chunk latency (LCL) (between the request being sent and the last chunk being received), and real-time factor RTF (LCL/DUR). For non-incremental synthesis (Non-INCR), we measure latency and RTF. Finally, We measure the mean opinion score (MOS) of the INCR, compared with the Non-INCR and the ground truth.

\subsection{Discussion}

\begin{table}[htbp]
\caption{Latency and RTF Comparison at 70 QPS}
\label{tab:latency_table}
\centering
\begin{tabular}{cccccc}
\toprule
TL&OL&DUR(s)&FCL(ms)&LCL(ms)&RTF\\
\midrule
Short&\multirow{4}{*}{4}&2.64&32.04&168.60&0.064\\
Med.& &4.08&38.16&295.36&0.072\\
Long& &7.00&77.05&1002.56&0.143\\
Mixed& &4.54&44.47&400.32&0.087\\
\midrule
Short&\multirow{4}{*}{8}&2.64&32.26&170.41&0.064\\
Med.& &4.08&39.38&305.88&0.075\\
Long& &7.00&91.42&1190.83&0.170\\
Mixed& &4.54&46.29&418.37&0.091\\
\midrule
Short&\multirow{4}{*}{-}&2.64&\multicolumn{2}{c}{227.66}&0.086\\
Med.& &4.08&\multicolumn{2}{c}{451.67}&0.109\\
Long& &7.00&\multicolumn{2}{c}{1405.51}&0.202\\
Mixed& &4.54&\multicolumn{2}{c}{1253.25}&0.313\\
\bottomrule
\end{tabular}
\end{table}

\footnotetext[1]{\href{https://muyangdu.github.io/Efficient-Incremental-TTS-on-GPUs}{https://muyangdu.github.io/Efficient-Incremental-TTS-on-GPUs}}

\subsubsection{Response Latency and Real-Time Factor}

We pressure-test the proposed approach using QPS (Queries-per-Second) from 10 to 100. Each test lasts 200 seconds. Requests sent to the server are evenly distributed within a second. Full performance metrics and samples can be found on our Github page\footnotemark[1]. Figure 3 shows the latency at different QPS and overlaps using mixed-length texts, which is more comparable to the production scenario. In the figure, we omit the data with a latency of more than 1s. For INCR, the first-chunk latency can be kept within 40 ms, and the last-chunk latency within 400 ms if the QPS does not exceed 60. In contrast, Non-INCR has a latency of more than 200 ms at 10 QPS and a latency of more than 800 ms at 60 QPS. The response latency of INCR has a 95.4\%, 91.2\% and 89.3\% deduction compare with Non-INCR at 60, 30, and 10 QPS, respectively. At 100 QPS, INCR can synthesize in real-time and provide a first-chunk response latency of less than 100ms, while Non-INCR is not real-time and its latency already far exceeds 1s. In regard to overlap, using 8 frames for overlap has slightly higher latency than 4 under high QPS. 

Figure 4 shows the variation of RTF with QPS up to 70, below which all the configurations can synthesize in real-time. INCR of short, medium, and mixed text maintains a smooth RTF below 0.1 within 70 QPS. In comparison, the RTF of Non-INCR on the mixed text has a dramatic increase after 50 QPS. This proves the effectiveness and stability of the proposed strategies. Table 1 shows the performance details at 70 QPS. We observe that the mixed LCL is comparable to the medium LCL for INCR. In contrast, the mixed latency is only comparable to the long latency for Non-INCR. Similar observation is also shown in figure 4, in which the mixed RTF is even greater than the long RTF for Non-INCR. We attribute this problem to the redundant padding introduced for batched inference of all the modules in Non-INCR. Although padding to the maximum sequence length within a batch is still required for the $F$ and $E$ in INCR, the redundant padding of the $D$ and $V$ has been greatly reduced, leading to more efficient synthesis on GPUs.

\begin{table}[htbp]
\caption{MOS with 95\% Confidence Intervals}
\label{tab:mos_score_table}
\centering
\begin{tabular}{ccc}
\toprule
Type& Overlap& MOS (CI)\\
\midrule
INCR&4& 4.213$ \pm $0.120\\
INCR&8& 4.237$ \pm $0.127\\
Non-INCR& -& 4.362$ \pm $0.143\\
\midrule
&Ground Truth& 4.625$ \pm $0.053\\
\bottomrule
\end{tabular}
\vspace{-5pt}
\end{table}

\subsubsection{Subjective Evaluation of Speech Quality}

Table 2 shows the MOS of incremental versus non-incremental synthesis and ground truth. The proposed approach can synthesize comparable speech quality with the non-incremental approach\footnotemark[1]. Moreover, we observe a weak trade-off between incremental speech quality and overlap length. Combined with the performance analysis in the previous subsection, we consider it worthwhile to use a larger overlap for a better speech quality.


\section{Conclusions}

In this paper, we propose a highly efficient approach for incremental text-to-speech synthesis on GPUs. The proposed approach can achieve ultra-low latency under high concurrency on a single GPU by applying Instant Request Pooling and Module-wise Dynamic Batching strategies. Experimental results and timeline analysis further prove the effectiveness of the proposal. Furthermore, our study provides a complete process incremental TTS performance benchmark on GPU. Future work will be focused on further improving efficiency and speech quality.

\bibliographystyle{IEEEtran}

\bibliography{mybib}


\end{document}